\documentclass[pra,superscriptaddress,showpacs,longbibliography,reprint]{revtex4-2} 

\usepackage[colorlinks=true,linkcolor=blue,urlcolor=blue,citecolor=blue,pdfusetitle]{hyperref}
\usepackage{color}
\usepackage[utf8]{inputenc}
\usepackage[usenames,dvipsnames]{xcolor}
\usepackage{amsmath}
\usepackage{amssymb}
\usepackage{graphicx}
\graphicspath{{graphics/}}
\usepackage{epsfig}
\usepackage{dcolumn}
\usepackage{bm}
\usepackage{mathrsfs}
\usepackage{multirow}
\usepackage[all]{xy}
\usepackage{pbox}
\usepackage{lipsum}
\usepackage{verbatim}
\usepackage{braket}
\usepackage{dsfont}
\usepackage{array}
\usepackage{makecell}
\usepackage{tabularx}
\usepackage{isotope}
\usepackage{xr}
\usepackage{float}

\definecolor{dgreen}{rgb}{0.0, 0.5, 0.0}

\newcommand{\df}{\mathrm{d}}

\newcommand{\Tr}{\mathrm{Tr}}

\newcommand{\tp}{\, .}

\begin{document}
\raggedbottom

\title{Nonequilibrium thermodynamics and power generation in open quantum optomechanical systems}

\newcommand{\tubingen}{Institut f\"{u}r Theoretische Physik, Eberhard Karls Universit\"{a}t T\"{u}bingen, Auf der Morgenstelle 14, 72076 T\"{u}bingen, Germany}

\author{Paulo J. Paulino}
\email{paulo.paulino.souza96@gmail.com}
\affiliation{\tubingen}

\author{Igor Lesanovsky}
\affiliation{\tubingen}
\affiliation{School of Physics and Astronomy and Centre for the Mathematics and Theoretical Physics of Quantum Non-Equilibrium Systems, The University of Nottingham, Nottingham, NG7 2RD, United Kingdom}

\author{Federico Carollo}
\affiliation{\tubingen}

\date{\today}

\begin{abstract}
Cavity optomechanical systems are a paradigmatic setting for the conversion of electromagnetic energy into mechanical work. 
Experiments with atoms coupled to cavity modes are realized in nonequilibrium conditions, described by phenomenological models encoding non-thermal dissipative dynamics and falling outside the framework of weak system-bath couplings. This fact makes their interpretation as quantum engines, e.g., the derivation of a well-defined efficiency, quite challenging.  
Here, we present a consistent thermodynamic description of open quantum cavity-atom systems. Our approach takes advantage of their nonequilibrium nature and arrives at an energetic balance which is fully interpretable in terms of persistent dissipated heat currents. The interaction between atoms and cavity modes can further give rise to nonequilibrium phase transitions and emergent behavior and allows to assess the impact of collective many-body phenomena on the engine operation. To enable this, we define two thermodynamic limits related to a weak and to a strong optomechanical coupling, respectively. We illustrate our ideas focussing on a time-crystal engine and discuss power generation, energy-conversion efficiency, and  emergence of metastable behavior in these limits. 
\end{abstract}

\maketitle

\section{Introduction}

\begin{figure}[t!]
    \centering
    \includegraphics[width=\columnwidth]{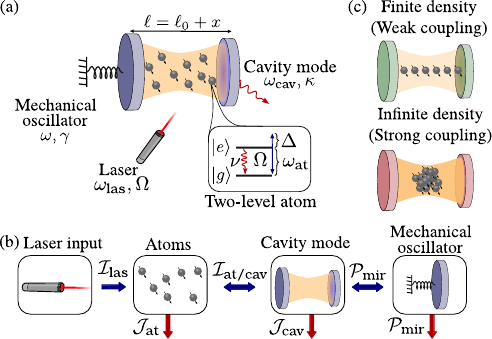}
    \caption{{\bf Nonequilibrium cavity-atom optomechanical engine.} (a) Atoms in a cavity are driven by a laser with Rabi-frequency $\Omega$ and detuning $\Delta$. One of the cavity mirrors can move allowing for oscillations of the cavity length $\ell$ from its equilibrium position $\ell_0$. (b) The laser provides energy which flows through the cavity to the mirror. Atom and photon losses at rates $\nu$ and $\kappa$, respectively, determine energy dissipation. The mechanical power delivered by the engine equals the power the mirror dissipates due to friction. (c) Two possible thermodynamic limits for the system. The first features a finite density of atoms and results in a weak optomechanical coupling. The second features an infinity density of atoms giving rise to a strong optomechanical coupling.}
    \label{fig:Fig1}
\end{figure}

The application of thermodynamics to quantum systems \cite{alicki1979,kosloff1984,quan2007,kosloff2013,binder2018,deffner2019quantum} allows to conceive quantum heat engines, which perform ideal cycles between thermal equilibrium states \cite{abah2012,rossnagel2016,bouton2021}.
In many experiments of interest, however, quantum systems are realized under genuine out-of-equilibrium conditions, for example in the case of experiments with cold atoms in optomechanical cavities \cite{brennecke2007,brennecke2008,murch2008,purdy2010,chen2011,dechiara2011,ritsch2013,stamper-kurn2014,labeyrie2014,mivehvar2021, mikaeili2023optomechanically} [see sketch in Fig.~\ref{fig:Fig1}(a)]. 
These systems absorb energy from an external source, e.g., a laser, which prevents them from equilibrating with their surrounding and gives rise to persistent energy currents. This aspect motivates the development of alternative nonequilibrium quantum-engine cycles \cite{elouard2018efficient,niedenzu2016}, with driving protocols that are not described by thermal dynamics \cite{feldmann2003,rezek2006}.  
It further poses the problem of devising theoretical approaches \cite{RevModPhys.93.035008,levy2014,barra2015,strasberg2017,dechiara2018,micadei2019reversing,Hewgill2021, https://doi.org/10.48550/arxiv.2211.12558} providing a \emph{consistent} thermodynamic understanding of established experimental models \cite{stockburger2017,dann2021,mivehvar2021,defenu2021}.
These open challenges do not solely concern quantum systems, but are of much broader relevance, as witnessed by recent efforts in characterizing work  in active matter \cite{solon2015pressure, PhysRevE.97.020602, solon2018generalized,fodor2022irreversibility, fodor2021active,PhysRevX.7.021007,PhysRevLett.117.038103, PhysRevLett.108.080402}.

In this paper, we focus on paradigmatic open quantum optomechanical systems which can nowadays be realized and efficiently controlled in experiments \cite{brennecke2007,brennecke2008,murch2008,purdy2010,chen2011,dechiara2011,ritsch2013,stamper-kurn2014,labeyrie2014,mivehvar2021,kippenberg2005analysis,jayich2008dispersive,chen2011,stamper-kurn2014,aspelmeyer2014cavity,PhysRevLett.129.063604,lei2022many}. These setups are promising for the conversion of electromagnetic  energy into mechanical work, both in equilibrium \cite{zhang2014,zhang2014b} and in nonequilibrium conditions \cite{elouard2015,kilmovsky2015,mari2015,brunelli2015}. 
For these systems, we develop a  thermodynamic description which  characterizes the power transferred by the cavity to the mechanical oscillator [cf.~Fig.~\ref{fig:Fig1}(b)] as well as the efficiency of such energy conversion. 
Our approach 
allows us to formulate an energy balance in terms of the persistent nonequilibrium heat currents [see Fig.~\ref{fig:Fig1}(b)]. 
To investigate the impact of phase transitions and collective behavior on the performance of nonequilibrium engines, we consider two different thermodynamic limits  \cite{reif2009fundamentals,minganti2018spectral, casteels2017critical, carmichael2015breakdown, carollo2020nonequilibrium,bibak2022}, see Fig.~\ref{fig:Fig1}(c). 
One features a weak optomechanical coupling, ensuing from a finite density of atoms in the cavity \cite{carollo2020nonequilibrium}, and is characterized by finite power and zero efficiency. 
The other one features a strong optomechanical coupling, due to an infinite density of atoms. In this case, the delivered mechanical power is extensive in the ``size" of the system and the efficiency is finite. 

We illustrate our ideas exploiting a time-crystal \cite{wilczek2012quantum,buca2019,iemini2018boundary,sacha2018,carollo2021} engine, which is a manifestation of a nonequilibrium many-body quantum engine \cite{PhysRevLett.111.070402, watanabe2015absence,carollo2020nonequilibrium}. Our results also apply to generic optomechanical settings \cite{aspelmeyer2014cavity,mivehvar2021} or related spin-boson models, such as Rydberg-atom systems with interacting electronic and vibrational degrees of freedom \cite{gambetta2020,carollo2020} or superconducting-qubit systems \cite{blais2021}.

\section{The model} 
We consider the setup in Fig.~\ref{fig:Fig1}(a), which shows an ensemble of $N$ atoms loaded into a cavity. The atoms are described by two-level systems with ground state $\ket{g}$, excited state $\ket{e}$ and bare Hamiltonian $H_\mathrm{at}=\hbar\omega_\mathrm{at}\sum_{k=1}^N n^{(k)}$, where $n=\ket{e}\!\bra{e}$.
The bare cavity Hamiltonian is $H_\mathrm{cav}=\hbar\omega_\mathrm{cav}a^\dagger a$, with $a,a^\dagger$ being the cavity-mode operators. The atoms and the cavity mode interact through a Tavis-Cummings term \cite{jaynes1963comparison, hepp1973superradiant, hepp1973equilibrium} 
\begin{equation}
H_\mathrm{int}=\frac{\hbar g}{\sqrt{N}}\left(aS_++a^\dagger S_-\right)\, ,
\end{equation}
where $S_-=\sum_{k=1}^N\sigma_-^{(k)}$, with $\sigma_-=\ket{g}\!\bra{e}$, and $S_+=S_-^\dagger$. The atoms are further driven by a laser which, in combination with the bare atom energy, is described by the Hamiltonian (in the interaction picture) \begin{equation}
H_{\mathrm{las}}=\hbar\Omega (S_-+S_+)-\hbar\Delta \sum_{k=1}^N n^{(k)}\, ,
\end{equation}
where $\Omega$ is the Rabi-frequency and $\Delta=\omega_\mathrm{las}-\omega_{\mathrm{at}}$ is the detuning of the laser frequency from the atom transition frequency [cf.~Fig.~\ref{fig:Fig1}(a)]. 

The dynamics of the system state $\rho_t$ is governed by the master equation \cite{breuer2002theory} (in the interaction picture) 
\begin{equation}
\dot{\rho}_t=\mathcal{L}[\rho_t]:=-\frac{i}{\hbar}[H,\rho_t]+\mathcal{D}_{\mathrm{at}} [\rho_t]+\mathcal{D}_{\mathrm{cav}} [\rho_t]\, , 
\label{generator}
\end{equation}
with Hamiltonian  $H=H_\mathrm{las}+H_\mathrm{int}-\hbar \delta a^\dagger a$ and $\delta=\omega_{\mathrm{las}}-\omega_{\mathrm{cav}}$. 
The dissipators $\mathcal{D}_{\mathrm{at/cav}}$ account for irreversible effects due to a coupling of the atoms and the light-field to thermal reservoirs at inverse temperature $\beta$. For the atoms, we have the dissipator 
\begin{equation*}
\begin{split}
\mathcal{D}_{\rm at}[\rho]=&\nu \sum_{k=1}^N \left(\sigma_-^{(k)}\rho \sigma_+^{(k)}-\frac{1}{2}\left\{n^{(k)},\rho\right\} \right)\\
&+\nu e^{-\beta \hbar  \omega_\mathrm{at}}\sum_{k=1}^N \left(\sigma_+^{(k)}\rho \sigma_-^{(k)}-\frac{1}{2}\left\{1-n^{(k)},\rho\right\} \right)\, ,
\end{split}
\label{D_at}
\end{equation*}
while for the light-field the dissipator reads  
\begin{equation*}
\begin{split}
\mathcal{D}_{\rm cav}[\rho]=&\kappa \left(a \rho a^\dagger -\frac{1}{2}\left\{a^\dagger a ,\rho\right\} \right)+\\
&+\kappa  e^{-\beta\hbar  \omega_\mathrm{cav}} \left(a^\dagger \rho a-\frac{1}{2}\left\{a a^\dagger ,\rho\right\} \right)\, .
\end{split}
\label{D_cav}
\end{equation*}
Both encode spontaneous atom (photon) decay at rate $\nu$ ($\kappa$) and atom (photon) excitation at rate $\nu e^{-\beta \omega_\mathrm{at}}$ ($\kappa e^{-\beta \omega_\mathrm{cav}}$).  The latter  excitation process is often irrelevant in experiments, since $\beta\hbar\omega_{\text{cav/at}}\gg 1$ \cite{mivehvar2021}.

The cavity further features a movable mirror [cf.~Fig.~\ref{fig:Fig1}(a)], with mass $m$, frequency $\omega$ and damping rate $\gamma$. For our purposes, we can consider it as a classical object, whose deviation $x_t$ from its equilibrium position $\ell_0$ evolves through the equation \cite{tome2015stochastic,seifert2012stochastic}  
\begin{equation}\label{eq:mirror}
    m\ddot{x}_t+ \gamma \dot{x}_t + m\omega^2 x_t = f_t\, .
\end{equation}
Here, $f_t=\hbar G\langle a^\dagger a\rangle_t$, with $G=\omega_\mathrm{cav}^0/\ell_0$, is the radiation-pressure force on the mirror  in the linear coupling regime, $|x_t|/\ell_0\ll1$, with $\omega_\mathrm{cav} \approx \omega_\mathrm{cav}^0(1 - x_t/\ell_0)$ \cite{marquardt2006dynamical,aspelmeyer2014cavity}. 

This system can be interpreted as an engine [cf.~Fig.~\ref{fig:Fig1}(b)], or more precisely as an optomechanical energy converter. Atoms and light-field represent the engine many-body {\it working fluid}. They absorb electromagnetic energy from the external driving and convert it into mechanical work which is delivered to the mirror to sustain its motion. The output power can be  estimated as the heat dissipated by the mirror due to friction, thereby modelling  a ``dissipative load" \cite{seah2018}.

\section{Nonequilibrium thermodynamics}

The mirror ``state” (solely specified by instantaneous position $x_t$ and  velocity $\dot{x}_t$) and the state of the cavity-atom system $\rho_t$ are in product form. Eq.~\eqref{generator} provides the reduced quantum-system dynamics, parametrically depending on $x_t$ via the cavity frequency $\omega_{\rm cav}$. Similarly, Eq.~\eqref{eq:mirror} provides the reduced mirror dynamics. This dynamical decoupling suggests that, also from a thermodynamic viewpoint, the cavity-atom system and the mirror can be regarded as uncorrelated. The mirror dynamics can be accounted for, e.g., within the framework of stochastic thermodynamics \cite{seifert2012stochastic}. The challenge is, however, to consistently characterize the cavity-atom quantum engine.

The master equation [cf.~Eq.~\eqref{generator}] is  ``local"   \cite{levy2014} as is obtained by a weak coupling of the system with a thermal bath \cite{breuer2002theory} which, however, solely considers the bare system Hamiltonians $H_{\rm at/cav}$. This can be seen by the fact that $\mathcal{D}_{\rm at/cav}$ do not implement transitions between eigenstates of the full Hamiltonian $H$ but rather of $H_{\rm at/cav}$. A thermodynamic approach considering as internal energy the expectation of total system Hamiltonian would thus run into inconsistencies \cite{barra2015,levy2014,stockburger2017}, since it generically predicts a negative entropy production for local master equations  \cite{brandner2016}. A textbook approach \cite{alicki1979,kosloff1984,quan2007,kosloff2013,binder2018,deffner2019quantum} considering the total Hamiltonian $H$ and deriving a ``global” master equation would not show any inconsistency. However, here we are interested in providing a consistent description for the experimentally-relevant \cite{ritsch2013,mivehvar2021} local master equation \eqref{generator}. 

\subsection{The first laws}

Our approach takes inspiration from the separation, done in the framework of stochastic thermodynamics \cite{tome2015stochastic,seifert2012stochastic}, between conservative forces [cf.~the harmonic force in Eq.~\eqref{eq:mirror}] and non-conservative external ones [cf.~the force $f_t$ in Eq.~\eqref{eq:mirror}]. This entails the definition of two internal energies, one for the atoms and one for the light-field, through the bare energies and the identification of the remainder of the Hamiltonian, e.g., the laser driving, as an external driving contribution.

We thus start defining the atom internal energy as $u_{t}^\mathrm{at}:=\langle H_\mathrm{at} \rangle_t$.
Then, considering the generator $\mathcal{L}^*$, which is the dual of the generator $\mathcal{L}$ acting on observables, we find 
\begin{equation}\label{eq:Energy_balance_at}
    \begin{split}
        \dot{u}_t^\text{at}  &= \Tr(H_\text{at}\mathcal{L}[\rho_{t}]) =\langle\mathcal{L}^*[H_\text{at}]\rangle_t\\
        &= \frac{i}{\hbar}\left<[H, H_\text{at}]\right>_t  +  \left<\mathcal{D}_\text{at}^*[H_\text{at}]\right>_t + \left<\mathcal{D}_\text{cav}^*[H_\text{at}]\right>_t \\
        &=\frac{i}{\hbar}\left<[H_\text{las}, H_\text{at}]\right>_t + \frac{i}{\hbar}\left<[H_\text{int}, H_\text{at}]\right>_t +  \left<\mathcal{D}_\text{at}^*[H_\text{at}]\right>_t  \, ,
    \end{split}
\end{equation}
where the dissipators
$\mathcal{D}^*_\text{at/cav}$ are the dual dissipators for $\mathcal{D}_\text{at/cav}$ in the Heisenberg picture. 
For the last equality, we used that $\mathcal{D}_{\rm cav}^*$ acts nontrivially only on light-field operators. 
From the last line in Eq.~\eqref{eq:Energy_balance_at}, we can already identify the physical meaning of the different terms. 
The first term describes how the internal energy of the atoms varies due to the laser driving.
The second describes the energy flows from the atoms to the light field, while the last term, which comes from the dissipator, describes the heat power exchanged by the atoms with their environment. 

Instead of looking at the instantaneous currents, we want to take an average over a time-window $\tau$.
The latter could be a period of the engine cycle or in general just a long time-window. 
Time-averaging Eq.~\eqref{eq:Energy_balance_at}, we find 
\begin{equation}
\label{sm_at_bal}
    \frac{1}{\tau}\int_0^\tau \df t \, \dot{u}_t^{\rm at}=\frac{u_\tau^{\rm at}-u_0^{\rm at}}{\tau}=\mathcal{I}_{\rm las}-\mathcal{I}_{\rm at}-\mathcal{J}_{\rm at}\, .
\end{equation}
Here, we have defined 
\begin{equation}
\begin{split}
\mathcal{I}_\mathrm{las}&=\frac{i}{\hbar\tau}\int_0^\tau \!\df t \, \langle [H_\mathrm{las},H_\mathrm{at}]\rangle _t\, ,\\
\mathcal{I}_\mathrm{at}&=-\frac{i}{\hbar\tau}\int_0^\tau \!\df t \, \langle [H_\mathrm{int},H_\mathrm{at}]\rangle _t\, , \\
\mathcal{J}_{\mathrm{at}}&=-\frac{1}{\tau}\int_0^\tau \! \df t  \, \langle \mathcal{D}^*_{\mathrm{at}}[H_{\mathrm{at}}]\rangle_t\, .
\end{split}
\label{fluxes}
\end{equation}
The term $\mathcal{I}_{\rm las}$ represents the  time-averaged input power that the atoms receives from the laser and the term $\mathcal{I}_{\rm at}$ is instead the average power exchanged between the atoms and the light field (in this convention it is positive when flowing from the atoms to the light-field). 
The third term, $\mathcal{J}_{\mathrm{at}}$, is the average heat power dissipated by the atoms into the environment (also this quantity is positive when energy is leaving the atoms). 
Assuming that the integration time $\tau$ is large and the internal energy does not grow indefinitely with time, we have that $\frac{u_\tau^{\rm at}-u_0^{\rm at}}{\tau}\to 0$ so that from the above relation we can write
\begin{equation}\label{eq:I_LAS}
\mathcal{I}_{\rm las}=\mathcal{I}_{\rm at } +\mathcal{J}_{\rm at}\, .
\end{equation}

The internal energy of the light-field is defined as $u_t^{\mathrm{cav}}:=\langle H_{\mathrm{cav}}\rangle_t$. By taking the time-derivative and following a procedure analogous to the one exploited in Eq.~\eqref{eq:Energy_balance_at}, we find
\begin{equation}
\dot{u}_t^{\mathrm{cav}}=\frac{i}{\hbar}\langle [H_{\mathrm{int}}, H_{\mathrm{cav}}]\rangle_t-f_t v_t +\langle \mathcal{D}_{\mathrm{cav}}^*[H_{\mathrm{cav}}]\rangle_t\, , 
\end{equation}
with $v_t:=\dot{x}_t$ being the mirror velocity. 
Taking the time-average over $\tau$, the power absorbed by the light-field due to the coupling with the atoms, $\mathcal{I}_{\mathrm{cav}}=i/(\hbar\tau)\int_0^\tau \df t \langle [H_{\mathrm{int}}, H_{\mathrm{cav}}]\rangle_t$, is
\begin{equation}\label{eq:I_CAV}   \mathcal{I}_{\mathrm{cav}}=\mathcal{P}_{\mathrm{mir}}+\mathcal{J}_{\mathrm{cav}} \; ,
\end{equation}
where
\begin{equation}
\begin{split}
\mathcal{P}_{\mathrm{mir}}&=\frac{1}{\tau}\int_0^\tau \!\df t f_t v_t \, , \\
\mathcal{J}_{\mathrm{cav}}&=-\frac{1}{\tau}\int_0^\tau \!\df t \langle \mathcal{D}_{\mathrm{cav}}^*[H_{\mathrm{cav}}]\rangle_t\, .
\end{split}
\label{J_cav-P_mir}
\end{equation}
The first term above is the average heat power exchanged by the light-field and the environment, while $\mathcal{P}_{\mathrm{mir}}$ is the power delivered by the cavity-atom system to the mirror. Exploiting the mirror internal energy $u_t^\mathrm{mir}=(mv_t^2 + m\omega^2x^2_t)/2$ \cite{tome2015stochastic,seifert2012stochastic} and Eq.~\eqref{eq:mirror}, we also find the relation  
\begin{equation}\mathcal{P}_{\mathrm{mir}}=\gamma/\tau \int_0^\tau \df t v_t^2 \; ,
\end{equation}
i.e, the power delivered by the cavity-atom system to the mirror is equal, over a long time window, to the power dissipated by the mirror due to friction [cf.~Fig.~\ref{fig:Fig1}(a-b)]. 

In order to find the total input power, we observe that the quantity $\mathcal{I}_{\rm cav}-\mathcal{I}_{\rm at}$ can be written as  
$$
\mathcal{I}_{\rm cav}-\mathcal{I}_{\rm at}=\frac{1}{\tau}\int_0^\tau \df t \,  \frac{i}{\hbar}\langle [H_{\rm int}, H_{\rm at}+H_{\rm cav}]\rangle_t\, .
$$
This quantity can be different from zero whenever the atoms and the light field are not on resonance, in which case it  represents an ``imbalance" between the power delivered by the atoms and the power absorbed by the light field.
We consider this imbalance, which comes from the interaction Hamiltonian and is due to energy gain or energy loss associated with the exchange of excitations, as an additional input power contribution. 
The rationale is that, in typical cavity-atom experiments, interactions between the atoms and the cavity field need to be ``facilitated" by means of an additional laser driving, for instance through stimulated Raman emissions \cite{kirton2019}. In this sense, the imbalance term $\mathcal{I}_{\rm cav}-\mathcal{I}_{\rm at}$ is analogous, in spirit, to the term $\mathcal{I}_{\rm las}$. 
The total input is thus $\mathcal{I}_{\rm in}=\mathcal{I}_{\rm las}+\mathcal{I}_{\rm cav}-\mathcal{I}_{\rm at}$, which combining Eqs.~\eqref{eq:I_LAS}-\eqref{eq:I_CAV}, can be written as
\begin{equation}\label{eq:input}
    \mathcal{I}_{\rm in}=\mathcal{J}_{\rm cav}+\mathcal{J}_{\rm at}+\mathcal{P}_{\rm mir} \tp 
\end{equation}

For the sake of simplicity, we have considered here a nonfluctuating mirror dynamics, which effectively accounts for a zero-temperature bath for the mirror. Similar results could be obtained for finite temperatures for the mirror. For the regimes investigated here, the power delivered by the engine would still be proportional to the square of the average mirror velocity, see, e.g.,  considerations in Ref.~\cite{carollo2020nonequilibrium}.

\subsection{The second law}

In order to formulate a consistent efficiency for the optomechanical  engine, we first need to show that the heat powers obey the inequality $\mathcal{J}_\mathrm{at}+\mathcal{J}_\mathrm{cav}\ge0$. This is achieved by proving a suitable second law of thermodynamics through a modification of Spohn's theorem~\cite{spohn1978,brandner2016,Hewgill2021}. 
To this end, we define the map $\mathcal{D}=\mathcal{D}_{\rm at}+\mathcal{D}_{\rm cav}$, whose stationary state is the thermal state $ \rho^\beta\propto e^{-\beta(H_{\rm at}+H_{\rm cav})}$. 
Next, we consider the von Neumann entropy of the quantum state $\rho_t$,  $S(\rho_t)=-\Tr\left[\rho_t \log \rho_t\right] $, and define the entropy production as 
\begin{equation}
    \begin{split}
        \sigma_t &=\dot{S}(\rho_t)-\beta\Tr\left[\mathcal{D}\left[\rho_t\right]\left(H_{\mathrm{at}}+H_{\mathrm{cav}}\right)\right]\, . 
    \end{split}
\end{equation}
In the above equation, the first and second terms are the entropy and heat variations, respectively, and we further note that $\dot{S}(\rho_t)=-\Tr\left[\mathcal{D}\left[\rho_t\right]\log \rho_t\right]$. 
The task  is now to show that the entropy production is always positive. We proceed by defining the relative entropy $S(\rho_t||\rho^\beta)=\Tr\left[\rho_t \left(\log \rho_t-\log \rho^\beta\right)\right]$, which can only decrease under the action of a completely positive trace-preserving map~\cite{RevModPhys.74.197}.
Thus 
\begin{equation}
    \frac{\df}{\df u}S(e^{u\mathcal{D}}[\rho_t]||e^{u\mathcal{D}}[\rho^\beta])\Big|_{u=0}\le 0 \tp
\end{equation}
Since $e^{u\mathcal{D}}[\rho^\beta]=\rho^\beta$, the derivative of the relative entropy evaluated in $u=0$ becomes 
\begin{multline}
        \frac{\df }{\df u}S(e^{u\mathcal{D}}[\rho_t]||e^{u\mathcal{D}}[\rho^\beta])\Big|_{u=0}=\\
        - \dot{S}(\rho_t)+\beta \Tr\left[\mathcal{D}[\rho_t](H_{\mathrm{at}}+H_{\mathrm{cav}})\right] \tp 
\end{multline}
The non-positivity of the above quantity then implies the non-negativity of the entropy production. 

Averaging the entropy production over a long time window $\tau$, we find
\begin{equation}
\frac{1}{\tau}\int_0^\tau \df t\,  \sigma_t=\frac{S(\rho_\tau)-S(\rho_0)}{\tau}+\beta (\mathcal{J}_\mathrm{at}+\mathcal{J}_\mathrm{cav})\ge0\, .
\end{equation}
Now, assuming that $S(\rho_t)$ does not grow indefinitely with time, the above implies  $\mathcal{J}_\mathrm{at}+\mathcal{J}_\mathrm{cav}\ge0$, which is the inequality that we need.  Before using this inequality for discussing the efficiency of the engine, we make an important remark on the above derivation. The main ingredient that we have exploited is that the reduced dynamics of the quantum state obeys the Lindblad equation Eq.~\eqref{generator}, which only parametrically depends on the instantaneous position of the mirror $x_t$. The reduced dynamics of the quantum system assumes this form since the state of the mirror, fully specified by $x_t$ and $\dot{x}_t$, and the quantum state are in a product form. 
The product structure of the quantum-classical state remains generically also when considering single trajectories of stochastic dynamical equations for the mirror. 
However, when the mirror is promoted to a quantum degree of freedom, the derivation above requires appropriate modifications, given the emergence of intrinsic quantum correlations among all subsystems.

\subsection{Efficiency}

We are now able to obtain a thermodynamically consistent efficiency $\eta$ of the energy conversion occurring in our optomechanical setup. The input power is equal, at long times, to the total power dissipated by the optomechanical system, i.e.,  $\mathcal{I}_\mathrm{in}=\mathcal{P}_{\mathrm{mir}}+\mathcal{J}_{\mathrm{cav}}+\mathcal{J}_{\mathrm{at}}$.
Thus, 
\begin{equation}
\eta=\frac{\mathcal{P}_{\mathrm{mir}}}{\mathcal{P}_{\mathrm{mir}}+\mathcal{J}_{\mathrm{cav}}+\mathcal{J}_{\mathrm{at}}}\le 1\, .
\label{eq:efficiency_general}
\end{equation}
The efficiency is bounded by one since both $\mathcal{P}_{\mathrm{mir}}$ and $\mathcal{J}_{\mathrm{at}}+\mathcal{J}_{\mathrm{cav}}$ are positive,  as shown in the previous subsection. 
Considering the total input power, Eq.~\eqref{eq:input}, we can express  $\mathcal{I}_\mathrm{in}$ as in the denominator of Eq.~\eqref{eq:efficiency_general} and identify a well-defined efficiency formulated in terms of the persistent heat currents. 

\section{Time-crystal engine} 

\begin{figure}[t]
    \centering
    \includegraphics[width=\linewidth,height=\textheight,keepaspectratio]{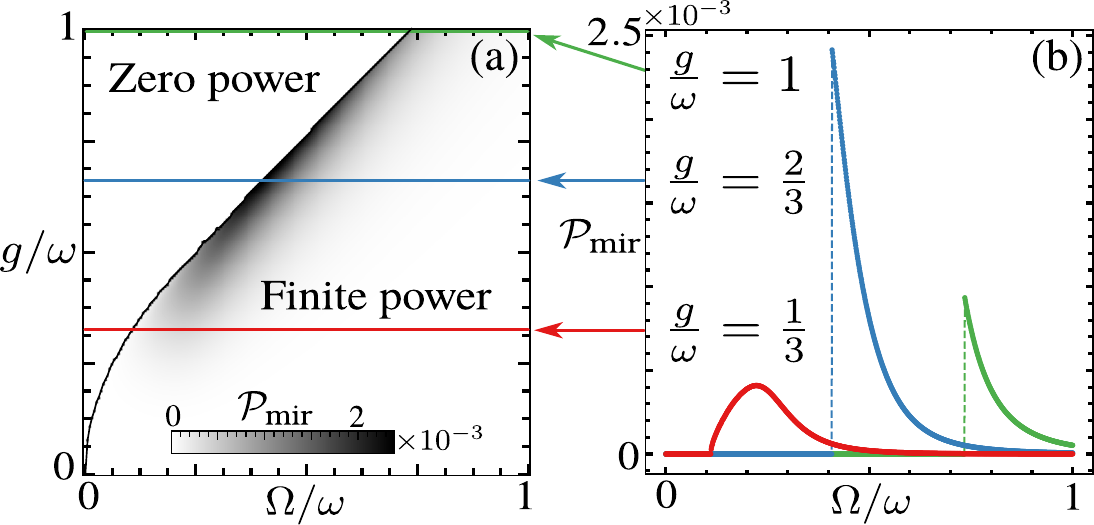}
    \caption{{\bf Mechanical power output of the time-crystal quantum engine.} (a) Power as a function of $\Omega$ and $g$, given in units of $\gamma (\Lambda\omega)^2$, with $\Lambda=\hbar G N/(m\omega^2)$. Note that, in our scaling limits, $\Lambda$ remains finite for all $N$. The parameter $\gamma$ is instead proportional to $N$ in the infinite-density limit in which the optomechanical engine thus delivers an extensive power output. The dotted line highlights the boundary between the stationary phase (zero power) and the time-crystal one (finite power). (b) Three sections across panel (a) for different values of $g$. The atoms are all initialized in the ground state and the light-field in the vacuum. The parameters used are $\kappa=\gamma/m=\omega$.}
    \label{fig:FDL}
\end{figure}

As an application of the general theory that we have developed, we analyze power and efficiency for the recently introduced time-crystal engine \cite{carollo2020nonequilibrium}.

For Markovian open quantum systems described by a time-independent Lindblad generator $\mathcal{L}$, one generically expects that the density matrix of the system  converges for long times to a stationary state, $\rho_{\infty}$, such that $\mathcal{L}[\rho_{\infty}]=0$. 
The generator $\mathcal{L}$ is time-translation invariant, due to its time-independence, and further commutes with the ``time-translation operator" (propagator) $e^{t\mathcal{L}}$.
Whenever the system approaches a steady-state, one has that $e^{t\mathcal{L}}[\rho_{\infty}]=\rho_{\infty}$, showing  that $\rho_{\infty}$ is {\it time-translation symmetric} and thus obeys the same symmetry of the generator (see also, e.g., the discussion in Ref.~\cite{carollo2021}).
The emergence of time-translation symmetry breaking, and thus of the so-called time-crystal phase, occurs when the state of the system approaches a limit cycle rather than a stationary state.
By denoting with $\rho^{\rm lc}_{t}$ the state inside such an asymptotic limit cycle and assuming that the latter has period $T$, we have that $\rho^{\rm lc}_{t+T}=\rho^{\rm lc}_{t}$. 
In such a case, we have that $e^{t'\mathcal{L}}[\rho^{\rm lc}_t]=\rho^{\rm lc}_{t+t'}\neq \rho^{\rm lc}_t$, whenever $t,t'\neq T$. This shows that, despite the fact that the generator $\mathcal{L}$ is time-translation symmetric, the asymptotic state of the system breaks such a symmetry. 
In this case, the system features persistent oscillations and is said to enter a time-crystal phase.
We note that signatures of the emergence of a  time-crystal phase can be seen in the spectrum of the (finite-system) generator $\mathcal{L}$ \cite{iemini2018boundary}.  

We now proceed by discussing how time-translation symmetry breaking can be used as a power-generation mechanism in our setup \cite{carollo2020nonequilibrium}.  In the regime in which the system approaches a stationary state, we have that the radiation-pressure force $f_t=\hbar \omega_{\rm cav}^0D_0|\alpha_t|^2$ becomes time-independent for long-times, since $\alpha_t$ approaches the stationary value associated with the stationary state.
In this case, the mirror is subject to a static force and thus its velocity converges to zero,   leading to a zero power production. 
On the other hand, in the time-crystal phase of the model, the state of the system features  persistent oscillations so that the radiation-pressure force $f_t$ remains asymptotically time-dependent.
As a consequence, the mirror is subject to a time-dependent force and thus never comes to rest and always sustains a finite velocity~\cite{iemini2018boundary,ritsch2013,mivehvar2021,tomadin2010,kirton2019,piazza2015}.
In the time-crystal regime, the mirror thus continuously dissipates power which must be provided by the time-crystal quantum engine. 
We note that, in the regime in which the Lindblad generator is time-independent, the engine is clearly in contact with a single bath at fixed temperature. As such, the optomechanical engine in this setup does not function as a heat engine but rather as a nonequilibrium isothermal one \cite{seifert2012stochastic}. 

\subsection{Mean-field treatment}

\begin{figure*}[t]
    \centering
    \includegraphics[width=\linewidth,height=\textheight,keepaspectratio]{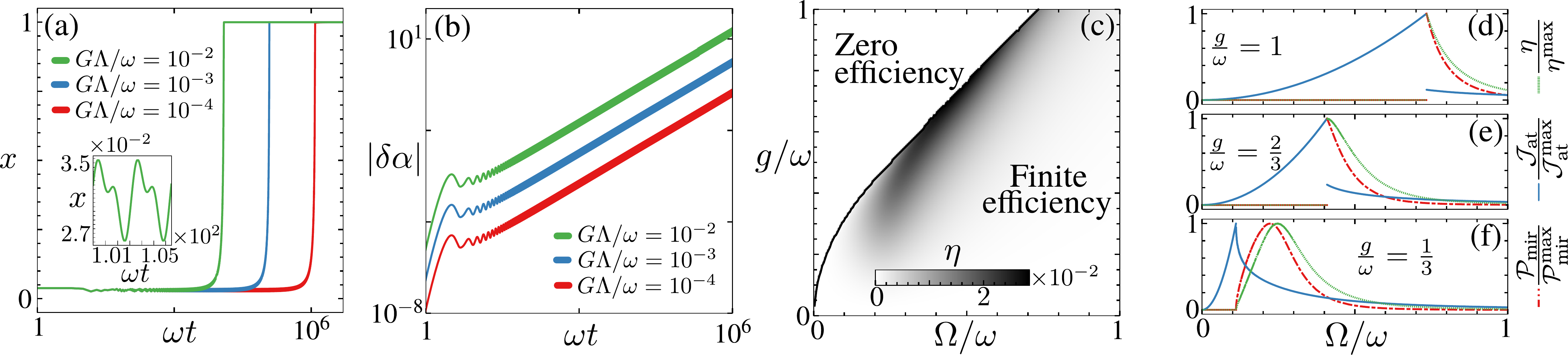}
\caption{{\bf Metastable time-crystal engine and efficiency in the infinite-density limit.}  (a) Mirror position $x$ in units of $\Lambda$ as a function of time for $\Omega=g=\omega$. The plot shows an emergent metastable regime in which the mirror features oscillations (see inset), unveiling that the quantum system is in a time-crystal phase. The smaller $G\Lambda/\omega$ the longer the metastable regime lasts, before the system ends up in the stationary state. (b) Growth of the fluctuation $\delta \alpha$, associated with the light-field expectation $\alpha$, due to the backaction of the dynamics of the mirror on the time evolution of the quantum system. (c) Efficiency --- in the metastable regime --- in units of $\gamma \Lambda/(\kappa m \ell_0)$, obtained from Eq.~\eqref{eq:efficiency}. (d-f) Normalized power, light-field dissipation and efficiency as a function of $\Omega$ for different values of $g$. The unspecified parameters are $\kappa=\gamma/m=\omega$. }\label{fig:IDL}
\end{figure*}

Since time-crystalline phases only emerge in the thermodynamic limit, we consider the system in the thermodynamic large $N$ limit, in which the quantum dynamics is exactly captured by a mean-field treatment \cite{carollo2021}.
That is, the rescaled operators $\alpha= a /\sqrt{N}$, $s_\pm= S_\pm/N$ and $s_z=S_z/N$, with $S_z=\sum_{k=1}^N(2n^{(k)}-1)$, converge, with $N\to\infty$, to scalar quantities evolving through nonlinear equations \cite{carollo2021,benatti2016non,benatti2018quantum}.
For concreteness, we consider the case $\omega_\text{at}-\omega_\text{cav}^0=\Delta=\nu=0$ and $\beta\to\infty$, for which the equations are given by
\begin{equation}
\begin{split}
\dot{s}_z&=-2\left[i\Omega s_++ig\alpha s_+ +\mathrm{c.c.}\right]\, ,\\
\dot{s}_+&=-i \left(\Omega+g\alpha^\dagger \right)s_z \, , \\
\dot{\alpha}&=-\frac{\kappa}{2}\alpha+i\left[Gx\alpha-g s_-\right]\, .
\end{split}
\label{mean-field}
\end{equation}
As shown below [see also Figs.~\ref{fig:FDL}-\ref{fig:IDL}], for large $\Omega/g$-ratios the system state approaches indeed  long-lived limit-cycle solutions \cite{iemini2018boundary,buca2019,carollo2022exact}.
In these cases, the radiation-pressure force is time-dependent and the mirror thus moves against friction, so that the cavity-atom engine delivers power [see $\mathcal{P}_{\mathrm{mir}}$ in Eq.~\eqref{J_cav-P_mir}] even without a time-dependent driving protocol \cite{carollo2020nonequilibrium}. 
In order to characterize the performance of the engine, we need to analyze the time evolution of the mirror and the dissipated power $\mathcal{P}_{\rm mir}$, in the thermodynamic limit. 
However, the force $f_t$ is extensive in $N$, $f_t\approx \hbar GN|\alpha_t|^2$, and this can give rise to an unphysical diverging displacement of the mirror.
To arrive at a well-defined mirror dynamics [see Eq.~\eqref{eq:mirror}], in the thermodynamic limit, we identify below two suitable scaling regimes, which are associated with two different physical scenarios. 

\subsection{ Finite-density (weak-coupling) limit}

First, we consider the regime in which, ideally, all atoms are located in the minima of an optical lattice inside the cavity [see sketch in Fig.~\ref{fig:Fig1}(c)]. The cavity length is thus proportional to the number of atoms \cite{carollo2020nonequilibrium,ritsch2013}, i.e., $\ell_0=N/D_0$, with $D_0$ being the linear density. The optomechanical coupling constant is here $G = \omega_\mathrm{cav}^0D_0/N$, and thus vanishes in the large $N$ limit. This scenario is associated with a weak optomechanical coupling [cf.~Fig.~\ref{fig:Fig1}(c)]. Looking at Eq.~\eqref{mean-field}, this  implies that the quantum-system dynamics does not depend  on any of the mirror parameters. Still, the mirror dynamics is driven by the light-field intensity through the force $f_t=\hbar \omega_\mathrm{cav}^0D_0|\alpha_t|^2$, which, in this regime, becomes independent on $N$. 

In Fig.~\ref{fig:FDL}(a), we show the power delivered by the engine to the mirror.
We observe a parameter region in which the delivered power $\mathcal{P}_{\mathrm{mir}}$ is zero. 
This occurs when the quantum system approaches a stationary state, the radiation-pressure force a stationary value, and the mirror comes to rest, as is expected for static driving.
Nonetheless, even with a static driving, for certain parameters the quantum system can spontaneously enter a state of sustained oscillations, determining a time-dependent force on the mirror and thus a finite power $\mathcal{P}_{\mathrm{mir}}$, see Fig.~\ref{fig:FDL}(b).
In this regime the optomechanical setup operates as a time-crystal quantum engine \cite{carollo2020nonequilibrium}.
However, in the finite-density limit  $\mathcal{P}_{\mathrm{mir}}$ is intensive in $N$, while $\mathcal{J}_{\mathrm{cav}}\propto N$, so that the engine efficiency is zero. \\

\subsection{ Infinite-density (strong-coupling) limit}

We now introduce a regime in which the cavity-atom engine operates with finite efficiency. 
We consider the limit $N\to\infty$ keeping $\ell_0$ finite, which leads to  an infinite density of atoms in the cavity [see sketch in Fig.~\ref{fig:Fig1}(c)] and to a finite $G$.  We dub this limit the ``strong" optomechanical coupling regime, since the force $f_t$ remains proportional to $N$.
To have a physically meaningful mirror dynamics [cf.~Eq.~\eqref{eq:mirror}],  we rescale the mass and the friction parameter as $m=N\tilde{m}$ and $\gamma =N \tilde{\gamma}$, respectively.
This means relating the ``size" of the mirror to the number of atoms, which is natural when the ``mirror" is a vibrational degree of freedom of the atom ensemble \cite{brennecke2008,murch2008,purdy2010, mikaeili2023optomechanically} or when the cavity hosts a cloud of atoms, as illustrated in Fig.~\ref{fig:Fig1}(c). 
In this way, the mirror velocity remains finite, while the power is extensive in $N$, since $\gamma\propto N$. 
This gives an efficiency (at lowest order in $\Lambda/\ell_0$)
\begin{equation}\label{eq:efficiency}
    \eta \approx   \frac{\gamma}{\kappa m}\frac{\Lambda}{\ell_0}\frac{\int_0^{ \omega \tau}{ d} [\omega t] V^2_{ \omega t}}{\int_0^{\omega \tau}{ d} [\omega t]|\alpha_{ \omega t}|^2 \, }\, . 
\end{equation}
Here, $\Lambda=\hbar G N/(m\omega^2) $ is the characteristic length-scale of $x$, while $V_{\omega t}$ is the dimensionless velocity at the dimensionless time $\omega t$, such that $v_t=\Lambda\omega V_{\omega t}$. 
The complete expression of the efficiency, including finite temperature and finite atom decay, is given in  Appendix~\ref{ap:Second_order}.

In this  infinite-density limit, the dynamics of the quantum system is influenced by the motion of the mirror [cf.~Eq.~\eqref{mean-field}]. 
For a time-scale of the order of $(G\Lambda)^{-1}$, this backaction is irrelevant and the cavity-atom system can host a (metastable) time-crystal phase, see Fig.~\ref{fig:IDL}(a-b), the Appendix~\ref{ap:metastable} and~\ref{ap:matrix_perturbation}.
Thereafter, backaction effects become non-negligible and drive the system towards a stationary state, where no power can be generated anymore. 

Even if it appears as a metastable phase, we can characterize the time-crystal engine in the long prestationary regime. The average power  is the same as that shown in Fig.~\ref{fig:FDL},  albeit being now extensive with $N$. The efficiency is reported in Fig.~\ref{fig:IDL}(c-d).
It also signals the transition from the stationary to the (metastable) time-crystal phase, where the system generates mechanical power. In Fig.~\ref{fig:IDL}(d), we see that for large $g/\omega$ the maximal power delivered by the engine occurs close to the transition line, where also a maximal efficiency is reached.  

\section{Discussion}

We characterized the nonequilibrium thermodynamics of  phenomenological cavity-atom  models. Contrary to other thermodynamic frameworks, our approach does not require the introduction of a repeated-interaction scheme \cite{barra2015,strasberg2017,dechiara2018}. It also does not consider as internal energy the expectation value of the total Hamiltonian \cite{kosloff2013,brandner2016,Hewgill2021}. We note that this is also what one would expected in the case of weak cavity-atom coupling, i.e., $\omega_{\rm at/cav}\gg |\Omega|,|g|$. Instead, it relies on separating, within the system Hamiltonian, the bare-energy contributions from those related to an external driving. This allows us to interpret dynamical contributions such as the laser driving as external power sources, which is closer, in spirit, to the physics of experiments with driven-dissipative quantum systems. Importantly, our identification of the different thermodynamic contributions leads to heat currents which are supported by a second-law of thermodynamics. This leads to a well-defined energy-conversion efficiency. 
For concreteness, we illustrated our ideas considering a time-crystal engine \cite{carollo2020nonequilibrium}. However, our approach may be applied more generally, also to cavity-only open quantum systems and to different manifestations of collective behavior hosted by them \cite{murch2008observation,maldovan2006simultaneous, eichenfield2009picogram,bibak2022,PhysRevLett.88.143601, frisk2019ultrastrong, PhysRevLett.128.207401}.

\section*{Acknowledgments} 
We are grateful to Albert Cabot and Stefano Marcantoni for useful discussions. We acknowledge funding from the Deutsche Forschungsgemeinschaft (DFG, German Research Foundation) under Project No. 435696605 and through the Research Unit FOR 5413/1, Grant No. 465199066. This project has also received funding from the European Union’s Horizon Europe research and innovation program under Grant Agreement No. 101046968 (BRISQ). F.C.~is indebted to the Baden-W\"urttemberg Stiftung for the financial support of this research project by the Eliteprogramme for Postdocs.

\appendix

\section{Details of the two thermodynamic limits} \label{ap:Second_order}
We provide here details on the discussion reported in the main text concerning the two thermodynamic limits. We start deriving the quantities of interest, considering a finite number of atoms $N$. However, we already introduce the mean-field approximation since the latter becomes exact in the limit $N\to\infty$. 

The time evolution of the relevant quantum operators is thus described by the mean-field equations reported in the main text. Through these operators we can write the heat currents as (derive for finite temperature)
\begin{equation}
    \begin{split}
    \mathcal{J}_{\mathrm{cav}}&=\frac{\hbar \kappa}{\tau}\int_0^\tau \df t \omega_\text{cav}\left<a^\dag a -e^{-\beta\hbar\omega_\text{cav}}(1+a^\dag a) \right> \;, \\ 
    \mathcal{J}_{\mathrm{at}}&=\frac{\hbar \omega_\text{at}\nu}{\tau}\int_0^\tau \df t \sum_{k=1}^N\left[\left(1+e^{-\beta\hbar \omega_\text{at}}\right)\langle n^{(k)}\rangle - e^{-\beta \hbar \omega_\text{at}}\right]
    \label{heat-fluxes}
    \end{split}
\end{equation}

To derive the mechanical output power, we need to look at the dynamics of the mirror, described by the second-order differential equation reported in the main text.
The solution to this equation is given by $x_t=\Lambda X_{\omega t}$, with $\Lambda=\frac{\hbar G N}{m \omega^2}$ and
\begin{equation}
    \begin{split}
X_{\omega t}=\int_0^{\omega t}{d[\omega s]} \, |\alpha_{\omega s}|^2 e^{-\frac{\gamma_0}{\omega}(\omega t -\omega s)}\frac{\omega}{\Sigma}\sin \left[\frac{\Sigma}{\omega} (\omega t -\omega s) \right]\, .
\label{position}
    \end{split}
\end{equation}
Here, $\gamma_0=\gamma/(2m)$ and we considered the dimensionless time $\omega t$. Moreover, we a slight abuse of notation we denoted with  $\alpha_{\omega t}$ the light-field operator at the dimensionless time $\omega t$. Furthermore, by taking the derivative, we find the velocity of the mirror as $v_t=\Lambda \omega V_{\omega t}$, with 
\begin{equation}
    \begin{split}
V_{\omega t}&=\int_0^{\omega t}{d[\omega s]} \, |\alpha_{\omega s}|^2 e^{-\frac{\gamma_0}{\omega}(\omega t -\omega s)}\\
&\times \left(\cos \left[\frac{\Sigma}{\omega} (\omega t -\omega s) \right]-\frac{\gamma_0}{\Sigma}\sin \left[\frac{\Sigma}{\omega} (\omega t -\omega s) \right]\right)\, . 
    \end{split}
    \label{velocity}
\end{equation}
Here, $V_{\omega t}$ is the dimensionless velocity at the dimensionless time $\omega t$. We note that $\dot{X}_{\omega t}=V_{\omega t}$ and $\dot{V}_{\omega t}=-X_{\omega t}-2\gamma_0/\omega V_{\omega t}+|\alpha_{\omega t}|^2$, which is a dimensionless system of equations. Solving this coupled with the mean-field equations, it is possible to compute the average mechanical output power as 
$$
\mathcal{P}_{\mathrm{mirr}}= \frac{\gamma}{\tau} \int_0^\tau {\df t} \, v_t^2=\gamma (\Lambda \omega)^2\left[\frac{1}{\omega \tau}\int_0^{\omega \tau} d[\omega t] V_{\omega t}^2\right]
$$

\subsection{Finite-density limit}
We start considering the finite-density limit. In this situation we have $G=\omega_{\mathrm{cav}}^0D_0/N$, which thus tends to zero in the large $N$ limit. This means that the mean-field equations become independent from $x_t$ and that the quantum system does not feel the back action due to the motion of the mirror. 

In the finite-density limit, we have that $\Lambda=\hbar \omega_{\mathrm{cav}}^0D_0/(m\omega^2)$ and that the power $\mathcal{P}_{\mathrm{mirr}}$ remains finite. However, as shown by Eqs.~\eqref{heat-fluxes}, the heat fluxes are extensive with $N$ so that the efficiency in this regime vanishes.

\subsection{Infinite-density limit}

In the infinite-density limit, the length of the cavity $\ell_0$ remains independent from $N$. This means that $G$ is finite in the thermodynamic limit. To have a well-defined dynamics for $x_t$ we thus rescale the mass of the mirror,  $m=N\tilde{m}$, as well as the friction parameter, $\gamma=N\tilde{\gamma}$. We note that the Eqs.~\eqref{position}-\eqref{velocity} remain valid and also that $\Lambda=\hbar G/(\tilde{m}\omega^2)$ does depend on $N$. 
The average mechanical power delivered by the optomechanical engine is now extensive, since 
$$
\mathcal{P}_{\mathrm{mirr}}= N\tilde{\gamma} (\Lambda\omega)^2 \left[\frac{1}{\omega \tau}\int_0^{\omega \tau} {d[\omega t]}  V_{\omega t}^2\right]\, .
$$
The efficiency can be computed as 
\begin{equation*}
    \eta=\frac{\mathcal{P}_{\mathrm{mirr}}}{\mathcal{J}_{\mathrm{cav}}+\mathcal{J}_{\mathrm{at}}+\mathcal{P}_{\mathrm{mirr}}}=\left(1+\frac{\mathcal{J}_{\mathrm{cav}}+\mathcal{J}_{\mathrm{at}}}{\mathcal{P}_{\mathrm{mirr}}}\right)^{-1}\, ,
\end{equation*}
and substituting for the relevant quantities we find 
\begin{equation}
\begin{split}
\eta&=\Bigg(1+\frac{\hbar \omega_{\mathrm{at}}\nu}{\tilde{\gamma}(\Lambda\omega)^2}\frac{\int_0^{\omega \tau}{d}[\omega t][(1+e^{-\beta \hbar \omega_{\mathrm{at}}})n_{\omega t}-e^{-\beta \omega_{\mathrm{at}}}]}{\int_0^{\omega \tau} {d[\omega t]} V_{\omega t}^2}+ \\
&+\frac{\hbar \omega_{\mathrm{cav}}^0\kappa }{\tilde{\gamma}(\Lambda \omega )^2}\frac{\int_0^{\omega \tau} {d[\omega t]}(1-\frac{\Lambda}{\ell_0}X_{\omega t})|\alpha_{\omega t}|^2(1-e^{-\beta \hbar \omega_\text{cav}})  }{\int_0^{\omega\tau} {d[\omega t]} V_{\omega t}^2}\Bigg)^{-1}\, ,
\end{split}
    \label{efficiency-global}
\end{equation}
where $n_{\omega t}$ is here the expectation value of the operator $n$ at the dimensionless time $\omega t$.

Specializing for the case $\nu=0$ and $\beta\to\infty$, we find after manipulating the parameters
\begin{equation}
 \eta \approx   \frac{\gamma}{\kappa m}\frac{\Lambda}{\ell_0}\frac{\int_0^{ \omega \tau}{ d} [\omega t] V^2_{ \omega t}}{\int_0^{\omega \tau}{ d} [\omega t]|\alpha_{ \omega t}|^2 \, }\, , 
    \label{eff_text}
\end{equation}
where we only consider the lowest order in $\Lambda /\ell_0$.

\section{Metastable time-crystal regime} \label{ap:metastable}
We provide here details on the emergence of a metastable time-scale where the optomechanical system works as a time-crystal engine with finite efficiency in the infinite-density limit. This time-scale emerges when considering the parameter $G\Lambda/\omega$ to be small as we now show. 

The evolution equation for $\alpha_{\omega t}$ is given by 
$$
\dot{\alpha}_{\omega t}=-\frac{\kappa}{2\omega}\alpha_{\omega t}+i\frac{G \Lambda}{\omega}\alpha_{\omega t}-i\frac{g}{\omega} s_{-\, \omega t}\, .
$$
Considering $G\Lambda/\omega$ small, we can apply perturbation theory on the mean-field equations. By doing this, one can show that (see next section for details)
\begin{equation}
\alpha_{\omega t}=\alpha^0_{\omega t}+\frac{G\Lambda}{\omega}\delta \alpha_{\omega t}^1\, ,
\label{perturbation-theory}
\end{equation}
where $\alpha^0_{\omega t}$ is obtained through the unperturbed system of mean-field equations, while the term $\delta \alpha_{\omega t}$ is the perturbation around this solution due to considering a small $G\Lambda/\omega$. In the plot in the main text, we provide the value $\delta \alpha=G\Lambda/\omega \delta\alpha^1$. 

Now we proceed by rewriting the second-order differential equation for the mirror oscillations in terms of the dimensionless quantity $\epsilon=Gx/\omega$. One readily obtains 
$$
\ddot{\epsilon}_{\omega t}+\frac{\gamma}{m\omega}\dot{\epsilon}_{\omega t}+\epsilon_{\omega t }=\frac{G\Lambda}{\omega}|\alpha_{\omega t}|^2\, .
$$
By recalling Eq.~\eqref{perturbation-theory}, we see that up to first-order in $G\Lambda/\omega$ the above equation is fully determined solely by the term $\alpha^0_{\omega t}$ which is given by the same system of equation solved for the finite-density limit and which can show persistent oscillations. 
However, the correction $\delta \alpha_{\omega t}$ increases linearly with time and thus eventually plays an important role in the dynamics of the mirror. Our exact numerical results show that the perturbation is such that the system will asymptotically approach a stationary state.

\section{Perturbation theory on the mean-field equations}\label{ap:matrix_perturbation} 
We give here a brief discussion on how we performed the perturbation theory to first-order in $G\Lambda/\omega$. 

In the dimensionless time, the mean-field equations become (we omit the explicit time-dependence)
\begin{equation}
\begin{split}
\dot{s}_z&=-2\left[i\frac{\Omega}{\omega} s_++i\frac{g}{\omega}\alpha s_+ +\mathrm{c.c.}\right]\, ,\\
\dot{s}_+&=-i \left(\frac{\Omega}{\omega}+\frac{g}{\omega}\alpha^\dagger \right)s_z\, ,\\
\dot{\alpha}&=-\frac{\kappa}{2\omega}\alpha+i\left[\frac{G\Lambda}{\omega}X\alpha-\frac{g}{\omega} s_-\right]\, ,
\end{split}
\label{mean-field-pt}
\end{equation}
where we used that $x=\Lambda X$. This equation is coupled with the second-order differential [see discussion below Eq.~\eqref{velocity}]
$$
\ddot{X}+2\frac{\gamma_0}{\omega}\dot{X}+X=|\alpha|^2\, .
$$
To zeroth order, we simply solve the above system setting $G\Lambda/\omega\equiv0$ which gives us the solution $s_z^0$, $s_+^0$ and $\alpha^0$. 

To first-order, we expect the solutions to be given by 
\begin{equation*}
    \begin{split}
    s_z&=s_z^0+\frac{G\Lambda}{\omega}\delta s_z^1 \; ,\\
    s_+&=s_+^0+\frac{G\Lambda}{\omega}\delta s_+^1\, ,\\
    \alpha&=\alpha^0+\frac{G\Lambda}{\omega}\delta \alpha^1\, .
    \end{split}
\end{equation*}
Substituting into the above system of equations, we find 
\begin{equation}
\begin{split}
\dot{\delta s_z^1}&=-2\left[i\frac{\Omega}{\omega}\delta s_+^1+i\frac{g}{\omega}(\alpha^0\delta s_+^1+s_+^0\delta \alpha^1)+\mathrm{c.c.}\right]\\
\dot{\delta s_+^1}&=-i\frac{\Omega}{\omega}\delta s_z^1 -i\frac{g}{\omega}\left(\alpha^{0\, \dagger}\delta s_z^1+s_z^0\delta \alpha^{1\, \dagger}\right)\\
\dot{\delta \alpha^1}&=-\frac{\kappa}{2\omega}\delta \alpha^1+iX^0\alpha^0-i\frac{g}{\omega}\delta s_-^1\, .
\end{split}
\end{equation}
Solving these equations, combined with the mean-field ones for the unperturbed variables  
$s_z^0$, $s_+^0$ and $\alpha^0$, gives the behavior of $\delta \alpha=(G\Lambda/\omega)\delta \alpha^1$ reported in the main text. 

\bibliography{refs}

\end{document}